\documentclass[10pt,conference]{IEEEtran}
\IEEEoverridecommandlockouts

\makeatletter

\def\endthebibliography{%
	\def\@noitemerr{\@latex@warning{Empty `thebibliography' environment}}%
	\endlist
}
\makeatother

\usepackage{cite}
\usepackage{amsmath,amssymb,amsfonts}
\usepackage[ruled,vlined]{algorithm2e}    
\usepackage{setspace}
\usepackage{algorithmic}

\usepackage{graphicx}
\usepackage{textcomp}
\usepackage{xcolor}
\def\BibTeX{{\rm B\kern-.05em{\sc i\kern-.025em b}\kern-.08em
    T\kern-.1667em\lower.7ex\hbox{E}\kern-.125emX}}

\usepackage{subfigure}
\usepackage{setspace}
\usepackage{multirow}
\usepackage{comment}
\usepackage{tabularx}
\usepackage{booktabs}
\usepackage{caption}

\usepackage[left=0.625in,right=0.625in,top=0.72in,bottom=0.97in]{geometry}

\usepackage[colorinlistoftodos]{todonotes}

\newcommand{\vectr}[1]{\boldsymbol{\mathrm{#1}}}
\newcommand{\matr}[1]{\boldsymbol{\mathrm{#1}}}

\newcommand\norm[1]{\left\lVert#1\right\rVert}

\newcommand{\RomanNumeralCaps}[1]{\MakeUppercase{\romannumeral #1}}

\newcommand{\herm}{\mathsf{H}}
\newcommand{\transp}{\mathsf{T}}
\newcommand{\trp}{\mathsf{T}}

\newcommand{\complexset}[2]{ \mathbb{C}^{#1 \times #2}  }
\newcommand{\Y}{\matr{Y}}
\newcommand{\G}{\matr{G}}
\newcommand{\g}{\vectr{g}}
\newcommand{\W}{\matr{W}}

\newcommand{\bde}{\gls{bde}\xspace}

\newcolumntype{L}{>{\centering\arraybackslash}m{4.5cm}}
\newcolumntype{K}{>{\centering\arraybackslash}m{2cm}}
\newcolumntype{R}{>{\centering\arraybackslash}m{4.5cm}}

\usepackage{xspace} % https://ianproberts.com/notes/latex_glossary.html
\usepackage[acronym,nogroupskip,nonumberlist,nopostdot]{glossaries}

\newacronym{2d}{2D}{two-dimensional}
\newacronym{3d}{3D}{three-dimensional}
\newacronym{3gpp}{3GPP}{third-generation partnership project}
\newacronym{adcs}{ADC}{analog-to-digital converters}
\newacronym{am}{AM}{amplitude modulation}
\newacronym{ambc}{AmBC}{ambient BC}
\newacronym{ap}{AP}{access point}
\newacronym{ar}{AR}{augmented reality}
\newacronym{aoa}{AOA}{angle-of-arrival}
\newacronym{awgn}{AWGN}{additive white Gaussian noise}
\newacronym{bc}{BC}{backscatter communication}
\newacronym{bde}{BD}{backscatter device}
\newacronym{ber}{BER}{bit error rate}
\newacronym{bibc}{BiBC}{bistatic BC}
\newacronym{ce}{CE}{carrier emitter}
\newacronym{csi}{CSI}{channel state information}
\newacronym{dmimo_s}{D-MIMO}{distributed MIMO}
\newacronym{dmimo}{D-MIMO}{distributed multiple-input multiple-output}
\newacronym{doa}{DOA}{direction-of-arrival}
\newacronym{dli}{DLI}{direct link interference}
\newacronym{en}{EN}{energy neutral}
\newacronym{end}{END}{energy neutral device}
\newacronym{etsi}{ETSI}{European Telecommunications Standards Institute}
\newacronym{evd}{EVD}{eigenvalue decomposition}
\newacronym{gs}{GS}{grid search}
\newacronym{gsm}{GSM}{Global System for Mobile Communications}  
\newacronym{gna}{GNA}{Girvan-Newman algorithm}
\newacronym{glrt}{GLRT}{generalized log-likelihood ratio test}
\newacronym{iot}{IoT}{Internet-of-Things}
\newacronym{los}{LoS}{line-of-sight}
\newacronym{lti}{LTI}{linear time-invariant}
\newacronym{lsb}{LSB}{least significant bit}
\newacronym{mimo}{MIMO}{multiple-input multiple-output}
\newacronym{miso}{MISO}{multiple-input single-output}
\newacronym{mmwave}{mmWave}{millimeter wave}
\newacronym{mmse}{MMSE}{minimum mean square error}
\newacronym{map}{MAP}{maximum a posteriori probability}
\newacronym{mrc}{MRC}{maximum-ratio combining}
\newacronym{mrt}{MRT}{maximum-ratio transmission}
\newacronym{mobc}{MoBC}{monostatic BC}
\newacronym{np}{NP}{Neyman-Pearson}
\newacronym{nfc}{NFC}{near-field communication}
\newacronym{ofdm}{OFDM}{orthogonal frequency division multiplexing}
\newacronym{p1}{P1}{Phase \RomanNumeralCaps{1}}
\newacronym{p2}{P2}{Phase \RomanNumeralCaps{2}}
\newacronym{pl}{PL}{path loss}
\newacronym{pana}{PanA}{Panel A}
\newacronym{panb}{PanB}{Panel B}
\newacronym{pcsi}{PCSI}{perfect channel state information}
\newacronym{papr}{PAPR}{peak-to-average power ratio}
\newacronym{pg}{PG}{path gain}
\newacronym{Riss}{RIS}{Reconfigurable intelligent surfaces}
\newacronym{ris}{RIS}{reconfigurable intelligent surface}
\newacronym{riss}{RIS}{reconfigurable intelligent surfaces}
\newacronym{rf}{RF}{radio frequency}
\newacronym{rfid}{RFID}{radio frequency identification}
\newacronym{rms}{RMS}{root mean square}
\newacronym{rss}{RSS}{received signal strength}
\newacronym{rv}{RV}{random variable}
\newacronym{snr}{SNR}{signal-to-noise ratio}
\newacronym{sinr}{SINR}{signal-to-interference-plus-noise ratio}
\newacronym{siso}{SISO}{single-input single-output}
\newacronym{simo}{SIMO}{single-input multiple-output}
\newacronym{tdoa}{TDOA}{time-difference-of-arrival}
\newacronym{toa}{TOA}{time-of-arrival}
\newacronym{tdd}{TDD}{time division multiplexing}
\newacronym{pgd}{PGD}{projected gradient descent}
\newacronym{ue}{UE}{user equipment}
\newacronym{uhf}{UHF}{ultra high frequency}
\newacronym{ula}{ULA}{uniform linear array}
\newacronym{upa}{UPA}{uniform planar array}
\newacronym{ura}{URA}{uniform rectangular array}
\newacronym{uwb}{UWB}{ultrawideband}
\newacronym{wpt}{WPT}{wireless power transfer}
\newacronym{zf}{ZF}{zero-forcing}
\newacronym{qam}{QAM}{quadrature amplitude modulation}

\newacronym{ofdma}{OFDMA}{orthogonal frequency-division multiple access}
\newacronym{fdd}{FDD}{frequency-division duplexing}
\newacronym{fdma}{FDMA}{frequency-division multiple access}
\newacronym{tdma}{TDMA}{time-division multiple access}
\newacronym{sdma}{SDMA}{space-division multiple access}
\newacronym{ls}{LS}{least-squares}
\newacronym{mse}{MSE}{mean square error}
\newacronym{fft}{FFT}{fast Fourier transform}
\newacronym{dft}{DFT}{discrete Fourier transform}
\newacronym{dtft}{DTFT}{discrete-time Fourier transform}
\newacronym{adc}{ADC}{analog-to-digital converter}
\newacronym{dac}{DAC}{digital-to-analog converter}
\newacronym{svd}{SVD}{singular value decomposition}
\newacronym{agc}{AGC}{automatic gain control}
\newacronym{nlos}{NLoS}{non-line-of-sight}
\newacronym{ple}{PLE}{path loss exponent}
\newacronym{eirp}{EIRP}{effective isotropic radiated power}
\newacronym{phy}{PHY}{physical layer}
\newacronym{4g}{4G}{fourth generation}
\newacronym{lte}{LTE}{Long-Term Evolution}
\newacronym{5g}{5G}{fifth generation}
\newacronym{nr}{NR}{New Radio}
\newacronym{5gnr}{5G NR}{5G New Radio}
\newacronym{ieee}{IEEE}{Institute of Electrical and Electronics Engineers}
\newacronym{lan}{LAN}{local area network}
\newacronym{wlan}{WLAN}{wireless local area network}
\newacronym{bs}{BS}{base station}
\newacronym{ul}{UL}{uplink}
\newacronym{dl}{DL}{downlink}
\newacronym{qos}{QoS}{Quality of Service}
\newacronym{sumimo}{SU-MIMO}{single-user \gls{mimo}}
\newacronym{mumimo}{MU-MIMO}{multi-user \gls{mimo}}
\begin{document}

\title{Access Point Selection for Bistatic Backscatter Communication in Cell-Free MIMO
	\thanks{©2024 IEEE. Personal use of this material is permitted. Permission from IEEE must be obtained for all other uses, in any current or future media, including reprinting/republishing this material for advertising or promotional	purposes, creating new collective works, for resale or redistribution to servers
	or lists, or reuse of any copyrighted component of this work in other works.
	}
}

\author{\IEEEauthorblockN{Ahmet Kaplan, Diana P. M. Osorio, and Erik G. Larsson}
	\IEEEauthorblockA{Department of Electrical Engineering (lSY), Linköping University, 581 83 Linköping, Sweden.\\
		Email: \{ahmet.kaplan, diana.moya.osorio, erik.g.larsson\}@liu.se}
}

\maketitle

\begin{abstract}
\Gls{bc} has emerged as a key technology to satisfy the increasing need for low-cost and green \gls{iot} connectivity, especially in large-scale deployments.
Unlike the \gls{mobc}, the \gls{bibc} has the possibility to decrease the round-trip path loss by having the \gls{ce} and the reader in different locations.  
Therefore, this work investigates the \gls{bibc} in the context of cell-free \gls{mimo} networks by exploring the optimal selection of CE and reader among all access points, leveraging prior knowledge about the area where the \bde is located. First, a \gls{map} detector to decode the \gls{bde} information bits is derived. Then, the exact probability of error for this detector is obtained. In addition, an algorithm to select the best \gls{ce}-reader pair for serving the specified area is proposed. Finally, simulation results show that the error performance of the \gls{bc} is improved by the proposed algorithm compared to the benchmark scenario. 
\vspace{-5pt}
\end{abstract}

\vspace{-5pt}
\glsresetall

\section{Introduction}
Passive \gls{iot}, a new paradigm based on battery-free devices, is a promising technology to enable several use cases that require connectivity with stringent requirements in terms of cost, complexity, and energy efficiency. These use cases span critical sectors, such as healthcare, transportation, and agriculture, thus the race toward the definition and standardization of passive \gls{iot} has already initiated, for instance, the \gls{3gpp} has run initial studies for the use cases and requirements in Release 18 \cite{3gpp.36.331, galappaththige2023cell}.
To that purpose, \gls{bc} has emerged as an enabling key technology for passive \gls{iot} by allowing devices to modulate its information on external \gls{rf} signals that are backscattered to the receiver or reader.

There are three types of \gls{bc}: \gls{mobc}, \gls{bibc}, and \gls{ambc}. 
In \gls{mobc}, the \gls{ce} and reader partially share the same hardware resources and are co-located. Therefore, \gls{mobc} requires full-duplex operation and suffers from the round-trip path loss effect \cite{kaplan2023direct}. 
\gls{mobc} has been investigated in  \cite{mishra2019optimal, kashyap2016feasibility, liu2014multi} by considering a system with multi-antenna technology to improve the communication range and throughput. 

In \gls{bibc}, the CE and reader are spatially separated from each other, thus the location of the CE and reader can be adjusted to decrease the round-trip path-loss effect \cite{kimionis2014increased}. In addition, \gls{bibc} operates in half-duplex mode which is less complicated compared to the full-duplex mode. In the literature, \gls{bibc} with multi-antenna technology is studied to improve the throughput, received SNR, and detection performance by direct link interference cancellation \cite{kaplan2023direct}, and optimal transmit and receive beamforming \cite{qu2022channel, galappaththige2023cell}.
Additionally, a channel estimation algorithm is proposed in \cite{rezaei2023time}.
Finally, the \gls{ambc} also consider that CE and reader are in different locations, while the CE is not considered dedicated, and uses ambient sources, such as Wi-Fi, Bluetooth, and TV signals, to transmit information \cite{van2018ambient, liu2013ambient, guo2018exploiting}. 

To circumvent the round-trip path loss effect, that degrades the communication quality and may prevent of providing sufficient power for the initial access of \bde \cite{galappaththige2023cell}, \gls{bibc} can explore the benefits of cell-free \gls{mimo} network deployment. Under this consideration, the best options for the \gls{ce} and reader in \gls{bibc} can be strategically selected to serve the \bde in order to decrease the round-trip path loss effect.

\Gls{ap} selection is an important and well-studied strategy in conventional cell-free \gls{mimo} systems to improve the overall performance \cite{chen2022survey,interdonato2019ubiquitous, ammar2021user}. 
In the context of cell-free architecture, a \gls{bibc} setup is investigated in \cite{galappaththige2023cell} by considering single antenna \glspl{ce} and a single reader with multiple antennas. Therein, optimal transmit and receive beamformers are designed to maximize the multiple \bde sum rate without considering the selection problem.

In this work, we focus on \gls{ap} selection in \gls{bibc} operating in a cell-free \gls{mimo} system. To the best of our knowledge, this is the first work that deals with AP selection strategy in this setup. Our contribution can be summarized as follows:
\begin{itemize}
	\item We derive the optimal \gls{map} detector to decode the information bits transmitted by the \bde operating in a \gls{bibc} setup within the cell-free \gls{mimo} network. We also derive a closed-form expression on the probability of error for the detector.
	\item Assuming that prior information about the area where a \bde is located is available, we show that selecting a single CE is the optimal scenario to serve the \gls{bde} located in the given area. 
	Under this consideration, we formulate and solve the problem of selecting the best CE among all \glspl{ap}.
	\item In addition, we propose a novel algorithm to select a single \gls{ce} and a single reader aiming to maximize the minimum quality of service for \gls{bc} in the given area.
	\item We show the superiority of the proposed algorithm over the benchmark in terms of probability of error.
\end{itemize}

\begin{figure}[h]
	\centering
	\includegraphics[width = 1\linewidth]{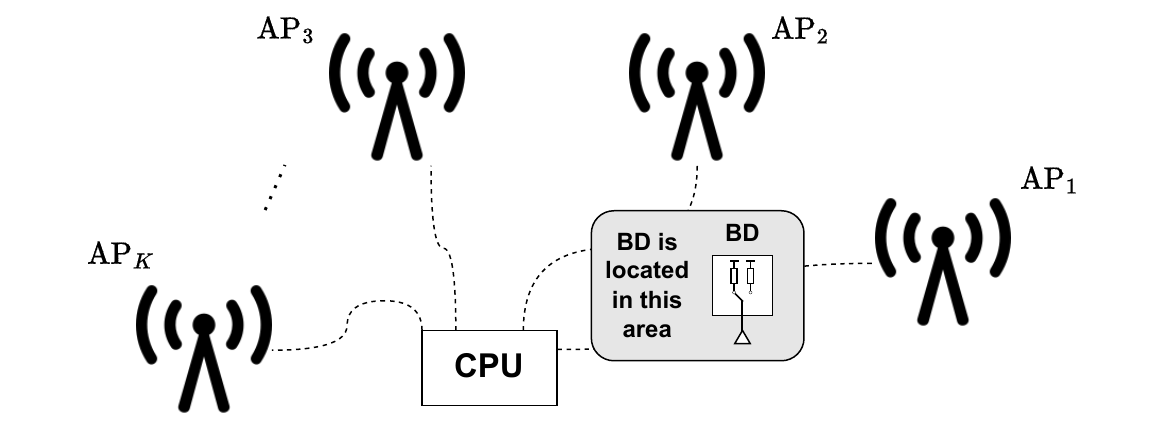}
	\vspace{-10pt}
	\caption{System model for the \gls{bibc} in a cell-free \gls{mimo} deployment.}
	\vspace{-15pt}
	\label{fig:System_Model}
\end{figure}
\vspace{-3pt}

\section{System Model}
\vspace{-5pt}
Consider the \gls{bibc} system illustrated in Fig. \ref{fig:System_Model}, which consists of
single-antenna \bde and $K$ \glspl{ap}, equipped with $M$ antennas, distributed across the coverage area. In this system, each AP is connected to a central processing unit through fronthaul links \cite{demir2021foundations}. The channel from \gls{ap} $t$ to \gls{ap} $r$ is given by $\matr{G}_{t,r} \in \complexset{M}{M}$, where $t$ and $r$ are elements of the set $\{1,2,\dotsc,K\}$. The channel from AP $r$ to AP $t$ is $\matr{G}_{t,r}^\trp$ due to reciprocity.
The channel from \gls{ap} $t$ to the \bde denoted by $\vectr{g}_t^\trp$, and the channel from the \bde to AP $t$ is $\vectr{g}_t \in \complexset{M}{1}$ due to reciprocity.

Herein, we assume that the exact location of \bde is unknown, but a region denoted as $\mathcal{B}$ where a \bde is located inside is given as prior information.
The precise location of the \bde can be unknown due to several reasons such as  noise and interference in the channel parameter estimation, insufficient energy in the \bde to send a pilot signal (especially before the initial access), and mobility of the \bde.
Therefore, the primary objective of our study is to define the best selection of one or more \glspl{ce} and readers among all possible APs to maximize the minimum quality of service for BDs located in the given region $\mathcal{B}$. 

\vspace{-3pt}
\section{Information Detection} \label{sec:info_det}
\vspace{-3pt}
In this section, we formulate the problem of information detection from the \bde. 
The null hypothesis $\mathcal{H}_{0}$ corresponds to sending bit $0$ and $\mathcal{H}_{1}$ corresponds to sending bit $1$ from the \bde. These hypotheses can be expressed as
\begin{equation} \label{eq:hypothesisTesting}
\vspace{-3pt}
	\begin{split}
		\mathcal{H}_{0}&:  \Y_{t,r} =\G_{t, r} \matr{\Phi} +\gamma_0 \g_{r} \g_{t}^\trp  \matr{\Phi}+ \W_{t,r},\\
		\mathcal{H}_{1}&: \Y_{t,r} =\G_{t, r} \matr{\Phi} +\gamma_1 \g_{r} \g_{t}^\trp  \matr{\Phi}+ \W_{t,r},
	\end{split}
\end{equation}
\vspace{-3pt}
\hspace{-5pt}where $\Y_{t,r}$$\in$ $\complexset{M}{\tau_d}$ is the received signal at AP $r$ when AP $t$ transmits. Here, $t \in \{t_{1},\ldots, t_{T}\}$ is the set of CEs and $r \in \{r_{1},\ldots, r_{R}\}$ is the set of readers. The matrix $\mathbf{W}_{t,r}~\in~\complexset{M}{\tau_d}$ denotes the additive Gaussian noise, and all elements of $\textbf{W}_{t,r}$ are independent and identically distributed (i.i.d.) $\mathcal{CN}(0, 1)$. 
The matrix $\matr{\Phi} \in \complexset{M}{\tau_d}$ denotes the orthogonal probing signal sent in a slot.
The probing signal $\matr{\Phi}$ satisfies $\matr{\Phi} \matr{\Phi}^{H} = \frac{p_t \tau_d}{M}\matr{I}_{M}$ and $\tau_d \geq M$, where $p_t$ stands for the transmit power and $\tau_d$ denotes the number of symbols in each time slot. The total amount of transmitted energy is expressed as $E_p  \triangleq||\boldsymbol{\Phi}||^2 = p_t \tau_d$. 
The reflection coefficient of the \bde is denoted as $\gamma \in \mathbb{R}$, and 
$\gamma=\gamma_0$ under $\mathcal{H}_{0}$ and $\gamma=\gamma_1$ under $\mathcal{H}_{1}$, where $\gamma_1>\gamma_0$.
We assume that the APs have \gls{pcsi}, i.e., $\G_{t,r}$, $\g_{r}$ and $\g_{t}$ are known. We also assume that $\matr{\Phi}$ is known.

Considering the system described above, the optimal \gls{map} detector to decode the \bde information can be expressed as
\begin{equation} \label{eq:map}
\widehat{\mathcal{H}}_{j}=\underset{\mathcal{H}_{j}}{\operatorname{argmax}}\ P(\mathcal{H}_{j} \mid \matr{Y}) =\underset{\mathcal{H}_{j}}{\operatorname{argmax}}\ p(\mathbf{Y} \mid \mathcal{H}_{j}) P(\mathcal{H}_{j}),
\end{equation}
where $\matr{Y}$ is a block matrix of the received signals, i.e., $\Y_{t,r}$.
The detector given in Eq. \eqref{eq:map} is equivalent to the following detector
\vspace{-8pt}
\begin{equation}\label{eq:NP}
	L(\matr{Y}) = \frac{ 
		\prod\limits_{t=t_1}^{t_T}\prod\limits_{\substack{r=r_1, r\neq t}}^{r_R}
		 p\left(\Y_{t,r} \mid \mathcal{H}_{1}\right)}
	{\prod\limits_{t=t_1}^{t_T}\prod\limits_{\substack{r=r_1, r\neq t}}^{r_R} p\left(\Y_{t,r} \mid \mathcal{H}_{0}\right)} \underset{\mathcal{H}_{0}}{\overset{\mathcal{H}_{1}}{\gtrless}} \frac{P(\mathcal{H}_{0})}{P(\mathcal{H}_{1})},
\end{equation}
where $P\left(\mathcal{H}_{0}\right)$ and $P\left(\mathcal{H}_{1}\right)$ are the prior probabilities. We assume that $P\left(\mathcal{H}_{0}\right)$$=$$P\left(\mathcal{H}_{1}\right)$$=$$1/2$.
The \gls{map} detector will decide $\mathcal{H}_{1}$, if the likelihood ratio, $L(\matr{Y})$, is bigger than the threshold. $p\left(\mathbf{Y}_{t,r} \mid \mathcal{H}_{1}\right)$ and $p\left(\mathbf{Y}_{t,r} \mid \mathcal{H}_{0}\right)$ denote the probability density functions (pdf) of the observations under $\mathcal{H}_{1}$ and $\mathcal{H}_{0}$, respectively, and are given by
\vspace{-6pt}
\begin{subequations} 
	\allowdisplaybreaks
	\begin{align}
		&p\left(\Y_{t,r} \mid \mathcal{H}_{1}\right)
			\nonumber \\
		&\qquad =\frac{1}{\pi^{M \tau_d}} \exp \left[-||\Y_{t,r}-\G_{t, r} \matr{\Phi} -\gamma_1 \g_{r} \g_{t}^\trp  \matr{\Phi}||^2\right], \\
		&p\left(\Y_{t,r} \mid \mathcal{H}_{0}\right) \nonumber \\
		&\qquad = \frac{1}{\pi^{M \tau_d}} \exp \left[-||\Y_{t,r}-\G_{t, r} \matr{\Phi} -\gamma_0 \g_{r} \g_{t}^\trp  \matr{\Phi}||^2\right].
	\end{align}
\end{subequations}
Let us define $\matr{A}_{t,r} \triangleq \g_{r} \g_{t}^\trp \matr{\Phi}$ and $\Y_{t,r}^\prime \triangleq \Y_{t,r} -\G_{t, r} \matr{\Phi}$. We can write $LLR=\log(L(\matr{Y}))$ as follows:

\begin{equation} 
	\begin{aligned}			
		LLR &= -\sum_{t}\sum_{r\neq t} ||\Y_{t,r}^\prime - \gamma_1\matr{A}_{t,r}||^2 
		   \\
		&\quad+ \sum_{t}\sum_{r\neq t} ||\Y_{t,r}^\prime - \gamma_0\matr{A}_{t,r}||^2.
	\end{aligned}
\end{equation}

Using the following equality 
	\begin{equation}
		\norm{\matr{B} - \matr{C}}^2=\norm{\matr{B}}^2 + \norm{\matr{C}}^2 - 2\operatorname{Re}\{\operatorname{Tr}\{\matr{C} \matr{B}^\herm\}\},
	\end{equation}
 one can show that the \gls{map} detector can be written as follows:
\begin{equation}
	\allowdisplaybreaks
	\begin{split}
	\label{eq:FinalGLRT}
	LLR^\prime &= \sum_{t}\sum_{r\neq t} \operatorname{Re}\{\operatorname{Tr}\{\matr{A}_{t,r} {\Y_{t,r}^\prime}^\herm\}\}  
	 \\
	&\underset{\mathcal{H}_{0}}{\overset{\mathcal{H}_{1}}{\gtrless}} \eta = \frac{(\gamma_1 + \gamma_0)}{2} \sum\limits_{t}\sum\limits_{r\neq t} ||\matr{A}_{t,r}||^2.
	\end{split}
\end{equation}
The distribution of the test statistic, $LLR^\prime$, both under $\mathcal{H}_{1}$ and $\mathcal{H}_{0}$ is given by
\begin{equation}
	LLR^\prime \sim \begin{cases}
		\mathcal{N}(\sum\limits_{t}\sum\limits_{r\neq t} \gamma_1 ||\matr{A}_{t,r}||^2, \frac{1}{2} \sum\limits_{t}\sum\limits_{r\neq t}||\matr{A}_{t,r}||^2) & \text {\hspace{-6pt}under } \mathcal{H}_{1} \\ 
		\mathcal{N}(\sum\limits_{t}\sum\limits_{r\neq t} \gamma_0 ||\matr{A}_{t,r}||^2, \frac{1}{2} \sum\limits_{t}\sum\limits_{r\neq t}||\matr{A}_{t,r}||^2) & \text {\hspace{-6pt}under } \mathcal{H}_{0}.
		\end{cases}
\end{equation}

The probability of error $(P_e)$ is given by
\begin{equation}	
	P_e = P(\mathcal{H}_{0} \mid \mathcal{H}_{1})P(\mathcal{H}_{1}) + P(\mathcal{H}_{1} \mid \mathcal{H}_{0})P(\mathcal{H}_{0}),
\end{equation}
and it is calculated as
\begin{equation}\label{eq:P_e_Case1_and_2}
	    P_e = Q\left((\gamma_1-\gamma_0)\sqrt{\frac{1}{2} \sum\limits_{t=t_1}^{t_T}\sum\limits_{\substack{r=r_1, r\neq t}}^{r_R} ||\matr{A}_{t,r}||^2} \right),
\end{equation}
where $Q(x)=\frac{1}{\sqrt{2 \pi}} \int_x^{\infty} \exp \left(-\frac{u^2}{2}\right) d u$, and $P_e$ decreases with the increasing of term $ \sum_{t}\sum_{r\neq t} ||\matr{A}_{t,r}||^2 = \frac{p_t \tau_d}{M} \sum_{t}\sum_{r\neq t} ||\g_{r}||^2||\g_{t}||^2$. 

\section{AP Selection}
In this section, we formulate the \gls{ap} selection problem. The channels are modeled as free-space \gls{los}, and the path-gain coefficients are defined as
\begin{equation}
		\beta_{t,(x,y)}= \frac{1}{d_{t,(x,y)}^2},
\end{equation}
where $d_{t,(x,y)}$ represents the distance between the center of AP $t$ and the \bde located at $(x,y)$ \cite[Sec. 7.2]{tse2005fundamentals}.
Therefore, the error performance of the detector depends on \gls{snr} expressed as
\begin{equation}
\sum\limits_{t=t_1}^{t_T}\sum\limits_{\substack{r=r_1 \\ r\neq t}}^{r_R} \frac{p_t \tau_d}{M} ||\g_{r}||^2||\g_{t}||^2 = \sum\limits_{t=t_1}^{t_T}\sum\limits_{\substack{r=r_1 \\ r\neq t}}^{r_R} \frac{p_t \tau_d M}{d_{r,(x,y)}^2 d_{t,(x,y)}^2}.
\end{equation}
We also define the transmit \gls{snr} as $p_t \tau_d$. Herein, it is assumed that all APs use the same transmit power, i.e., $p_t$.
We investigate two different cases for the AP selection problem, $2$ AP case and $2+$ AP case, which are detailed next. 

\subsection{$2$ AP Case} \label{sec:twoAP}
We assume that a region called $\mathcal{B}$ where a \bde is located inside is given as a prior information. If the $r$-th AP is selected as reader, the received signal at this AP is
\begin{equation}
	\matr{Y}_{t,r}= \matr{G}_{t,r}\matr{\Phi} + \gamma \vectr{g}_{r}\vectr{g}_{t}^\transp\matr{\Phi} + \matr{W}_{t,r}, 
\end{equation}
where $t,r \in \{1,2\}$, $t \neq r$, and $\gamma \in \{\gamma_0, \gamma_1\}$. 
The received \gls{snr} is expressed as
\begin{equation} \label{eq:snr1}
	\text{SNR} = \frac{p_t \tau_d}{M} \norm{\vectr{g}_r}^2 \norm{\vectr{g}_t}^2.
\end{equation}
 As seen in Eq. \eqref{eq:snr1}, the selected role of APs as CE and reader does not affect the SNR.

\subsection{2+ AP Case} 
As shown in Section \ref{sec:twoAP}, the AP role selection is trivial in the two-AP case. Therefore, in this subsection, we explore the use of a max-min fairness policy in setups involving more than two APs. 

We first prove that selecting a single AP as CE is the optimal approach to maximize the error performance when APs transmit orthogonal signals. For that, we investigate the error performance of three different cases. In all cases, the total radiated energy in each time slot is the same, irrespective of the number of \glspl{ce}, and there are a total of $T$ time slots available.

 \textbf{Case 1:} In the first case, assume that we have $K$ APs and $T$ among them are selected as CE. The set of selected CEs is  denoted as $t \in \{t_{1},\ldots, t_{T}\}$. It is assumed that there are $T$ time slots, and each CE sends a probing signal to the channel during its time slot, and the transmitted signal is received by $K$$-$$1$ APs (readers) in each time slot. 
 For instance, if $t \in \{2,3\}$, indicating two time slots, AP 2 transmits the probing signal in the first slot, which is received by the remaining APs. 
 Subsequently, in the second slot, AP 3 transmits its probing signal, which is then received by the other APs, including AP 2. This example is illustrated in Table \ref{tab:scenarious}. 
	Then, the detector coherently processes all the received signals, and the error performance of this scenario is directly  proportional to
	\begin{equation} \label{eq:lambda1}
	\vspace{-5pt}
		\Lambda_1 = \sum\limits_{t=t_1}^{t_T}\sum\limits_{\substack{r=1, r\neq t}}^{K} \frac{1}{d_{r,(x,y)}^2 d_{t,(x,y)}^2}.
		\vspace{-2pt}
	\end{equation}
	
	\textbf{Case 2:} In the second case, only the nearest AP to \bde, denoted as AP $t^\prime$, transmits the probing signal over $T$ time slots, while the remaining APs are assigned as readers, i.e., $r \in \{1,2,\dotsc,K\} \backslash t^\prime$.
	Consequently, the detection performance depends on the metric given by
	\begin{equation}
		\Lambda_2 = \frac{T}{d^2_{t^\prime,(x,y)}} \sum\limits_{\substack{r=1, r\neq t^\prime}}^{K} \frac{1}{d_{r,(x,y)}^2},
	\end{equation}
	where $d_{t^\prime,(x,y)}^2 = \underset{k}{\text{min}}(d_{k,(x,y)}^2), k = \{1,2,\dotsc,K\}$. It can be shown that 
	\begin{equation} \label{eq:lambda1vs2}
			\Lambda_1 \leq
			\sum\limits_{\substack{r=1, r\neq t^\prime}}^{K} \frac{T}{d_{r,(x,y)}^2 \underset{k}{\text{min}}(d_{k,(x,y)}^2)} = \Lambda_2.
	\end{equation}
	
	\begin{table}[tbp]
		\caption{An example of selected CEs, transmitted signals, and the number of readers for each time slot in all cases.}
		\vspace{-5pt}
		\centering
		\label{tab:scenarious}
		\resizebox{0.49\textwidth}{!}{%
			\begin{tabular}{|l|ccc|ccc|}
				\hline
				\multirow{2}{*}{}                                      & \multicolumn{3}{c|}{\textbf{Time Slot 1}}                                                       & \multicolumn{3}{c|}{\textbf{Time Slot 2}}                                                       \\ \cline{2-7} 
				& \multicolumn{1}{c|}{\textbf{CE}} & \multicolumn{1}{c|}{\textbf{Tr. Sig.}} & \textbf{Num. Read.} & \multicolumn{1}{c|}{\textbf{CE}} & \multicolumn{1}{c|}{\textbf{Tr. Sig.}} & \textbf{Num. Read.} \\ \hline
				\textbf{Case 1}                                        & \multicolumn{1}{c|}{AP $2$}        & \multicolumn{1}{c|}{$\matr{\Phi}$}                  & $K-1$                 & \multicolumn{1}{c|}{AP $3$}        & \multicolumn{1}{c|}{$\matr{\Phi}$}                  & $K-1$                 \\ \hline
				\textbf{Case 2}                                        & \multicolumn{1}{c|}{AP $t^\prime$}        & \multicolumn{1}{c|}{$\matr{\Phi}$}                  & $K-1$                  & \multicolumn{1}{c|}{AP $t^\prime$}        & \multicolumn{1}{c|}{$\matr{\Phi}$}                  & $K-1$                  \\ \hline
				\multicolumn{1}{|c|}{\multirow{2}{*}{\textbf{Case 3}}} & \multicolumn{1}{c|}{AP $2$}        & \multicolumn{1}{c|}{$\matr{\Phi}$}                  & $K-2$                  & \multicolumn{1}{c|}{AP $2$}        & \multicolumn{1}{c|}{$\matr{\Phi}$}                  & $K-2$                 \\ \cline{2-7} 
				\multicolumn{1}{|c|}{}                                 & \multicolumn{1}{c|}{AP $3$}        & \multicolumn{1}{c|}{$\matr{\Phi}$}                  & $K-2$                 & \multicolumn{1}{c|}{AP $3$}        & \multicolumn{1}{c|}{$-\matr{\Phi}$}                  & $K-2$                 \\ \hline
			\end{tabular}%
		}
		\vspace{-15pt}
	\end{table}
		
	\textbf{Case 3:} In cell-free \gls{mimo} systems, more than one AP can send their signals to the channel using the same time and frequency resources. Therefore, we also investigate the case that $T$ APs send orthogonal sequences of orthogonal signals into the channel over $T$ time slots simultaneously, and the number of readers in each time slot is $K-T$. The rows of Hadamard matrix, Zadoff-Chu sequence, and the rows of the \gls{dft} matrix can be used to design orthogonal sequences \cite{rezaei2023time}. For example, when $t \in \{2,3\}$, AP 2 and AP 3 are CEs, and $r \in \{1,2,\dotsc,K\} \backslash \{2,3\}$. AP 2 transmits $\matr{\Phi}$ in both time slots, and AP 3 transmits $\matr{\Phi}$ in the first time slot and $-\matr{\Phi}$ in the second time slot, as seen in Table \ref{tab:scenarious}. 
	It can be shown that the error performance is determined by $\Lambda_3$, given by
	\begin{equation}  
	\label{eq:lambda3}
		\Lambda_3 = \sum\limits_{t=t_1}^{t_T}\sum\limits_{\substack{r=1 \\ r\notin \{t_{1},\ldots, t_{T}\}}}^{K} \frac{1}{d_{r,(x,y)}^2 d_{t,(x,y)}^2} \leq \Lambda_2.
\end{equation}
In the derivation of $\Lambda_3$, we keep the total radiated energy in each time slot the same with the other cases for a fair comparison. 
\footnote{Please note that one might also have a per-AP power constraint, i.e., a per-AP energy constraint per slot. In this scenario, although Case 3 could be the most favorable option, it comes at the cost of radiating $T$ times more energy than the other two cases.} 
More details on the derivation are provided in Appendix \ref{FirstAppendix}.
Note that, in this case, the received signal is the sum of the transmitted signals from each CE.
Due to the use of orthogonal sequences in \glspl{ce}, all the received signals from different \glspl{ce} are separable in a reader. As a result, $\Lambda_1$ and $\Lambda_3$ are similar to each other except the indices of the summations. In Case 3, an AP either CE or reader and the set of readers is the same for all time slots, while in Case 1, each CE is also in the set of readers when they do not transmit. Therefore, for the same set of CEs, $\Lambda_1 > \Lambda_3$ when $T>1$ because there are $T-1$ more readers in Case 1 compared to Case 3.

In addition, as can be noticed from Eqs. \eqref{eq:lambda1}, \eqref{eq:lambda1vs2}, and \eqref{eq:lambda3}, $\Lambda_2 \geq \Lambda_1 \geq \Lambda_3$. As a result, selecting the best CE is the optimal strategy to minimize the error probability.
Therefore, we consider two scenarios for the AP selection:
(\romannumeral 1) the selection of a single CE while the remaining APs operate as readers, and (\romannumeral 2) the selection of one CE and one reader.

\subsubsection{Scenario 1 - Optimal CE Selection} \label{sec:optimalCE}
In this scenario, a single CE is chosen, while the remaining APs operate as readers in receiver mode.
When the $t$-th AP is selected as the CE, the received signal is represented as:
\begin{equation}
	\matr{Y}_{t,r}= \matr{G}_{t,r}\matr{\Phi} + \gamma \vectr{g}_{r}\vectr{g}_{t}^\transp\matr{\Phi} +  \matr{W}_{t,r}, 	
	\vspace{-5pt}
\end{equation}
for $r \in \{1,2,\dotsc, K\} \backslash t$. 

We use $\Lambda_2/T$ to investigate the optimal CE selection using the max-min fairness policy. The optimization problem can be formulated as
		\vspace{-4pt}
\begin{equation}
	\vspace{-3pt}
	\begin{aligned}
	   \mathcal{O} \mathcal{P}_{\text{C}}: \quad & \underset{t}{\text{max}}\underset{(x,y)}{\text{min}}
		& & \Lambda_2/T \\
		& \text{subject to}
		& & t \in \{1, \ldots, K\}, \\
		& & & (x,y) \in \mathcal{B},
	\end{aligned}
\end{equation}
where $(x,y)$ is a point within the region $\mathcal{B}$, where the BD is located inside. The problem $\mathcal{O} \mathcal{P}_{\text{C}}$ can be divided into two sub-problems. First, for a given $t$, we can formulate $\mathcal{O} \mathcal{P}_{\text{C}1}$ as
\begin{equation} \label{eq:minimization}
	\begin{aligned}
		\mathcal{O} \mathcal{P}_{\text{C}1}: \quad & \underset{(x,y)}{\text{min}}
		& & \frac{1}{d_{t,(x,y)}^2} \sum\limits_{\substack{r=1 \\ r\neq t}}^{K} \frac{1}{d_{r,(x,y)}^2} \\
		& \text{subject to}
		& &(x,y) \in \mathcal{B}. \\
	\end{aligned}
\end{equation}
Let us define the minimum value of $\mathcal{O} \mathcal{P}_{\text{C}1}$ for a given $t$ as $m_t$. Then, the second sub-problem, $\mathcal{O} \mathcal{P}_{\text{C}2}$, can be formulated as
\begin{equation} 
	\begin{aligned}
		\mathcal{O} \mathcal{P}_{\text{C}2}: \quad & \underset{t}{\text{argmax}}
		& & m_t \\
		& \text{subject to}
    	& & t = 1, \ldots, K.
	\end{aligned}
\end{equation}
The optimal $t$ for $\mathcal{O} \mathcal{P}_{\text{C}2}$ is the optimal CE and the solution for $\mathcal{O} \mathcal{P}_{\text{C}}$. The details of the solution for $\mathcal{O} \mathcal{P}_{\text{C}}$ and the sub-problems are provided in Section \ref{sec:algortihm_to_finf_optimal}.

\subsubsection{Scenario 2 - Optimal CE and Single Reader Selection}
In Scenario 1, the number of readers is $K-1$. In this scenario, we search for the optimal single reader along with the optimal single CE. 

The problem of selecting one CE and one reader is formulated as
\begin{equation}
	\begin{aligned}
		\mathcal{O} \mathcal{P}_{\text{CR}}: \quad & \underset{(t,r)}{\text{max}}\underset{(x,y)}{\text{min}}
		& & \frac{1}{d_{t,(x,y)}^2 d_{r,(x,y)}^2} \\
		& \text{subject to}
		& & t, r \in \{1, \ldots, K\}, \text{ and } t \neq r \\
		& & &(x,y) \in \mathcal{B}.
	\end{aligned}
\end{equation}
The details of the solution for $\mathcal{O} \mathcal{P}_{\text{C}}$ and $\mathcal{O} \mathcal{P}_{\text{CR}}$ are given in the next section.

\section{Proposed Methods to Find the Optimal CE and Optimal CE-Reader Pair} \label{sec:algortihm_to_finf_optimal}
\vspace{-2pt}

\subsection{Scenario 1 - Optimal CE Selection}
\vspace{-2pt}
We can rewrite Problem $\mathcal{O} \mathcal{P}_{\text{C}1}$ as follows:
\begin{equation}
	\begin{aligned}
		& \underset{(x,y)}{\text{min}}
		& & \frac{1}{(x_t-x)^2+(y_t-y)^2} \sum\limits_{\substack{r=1 \\ r\neq t}}^{K} \frac{1}{(x_r-x)^2+(y_r-y)^2} \\
	\end{aligned}
	\vspace{-5pt}
\end{equation} 
subject to $(x,y) \in \mathcal{B}$, where $x_t$ and $y_t$ are the coordinates of the position of $t$-th AP. To solve this problem, we apply \gls{pgd} for each AP $t$. Since the objective function is non-convex, we apply the \gls{pgd} method with multiple initial points. For that, we first partition the region $\mathcal{B}$ into small-sized areas, and we select the centroid of each area as an initial point in the PGD method. It is also possible to select multiple random initial points. Once we determine the minimum value of the objective function, i.e., $m_t$, we simply solve $\mathcal{O} \mathcal{P}_{\text{C}2}$ by comparing different values of $m_t$ to find the optimal CE.

\subsection{Scenario 2 - Optimal CE and Single Reader Selection} \label{sec:bestCEreaderPairs}
\begin{figure}[h]
	\centering
	\includegraphics[width = 0.55\linewidth]{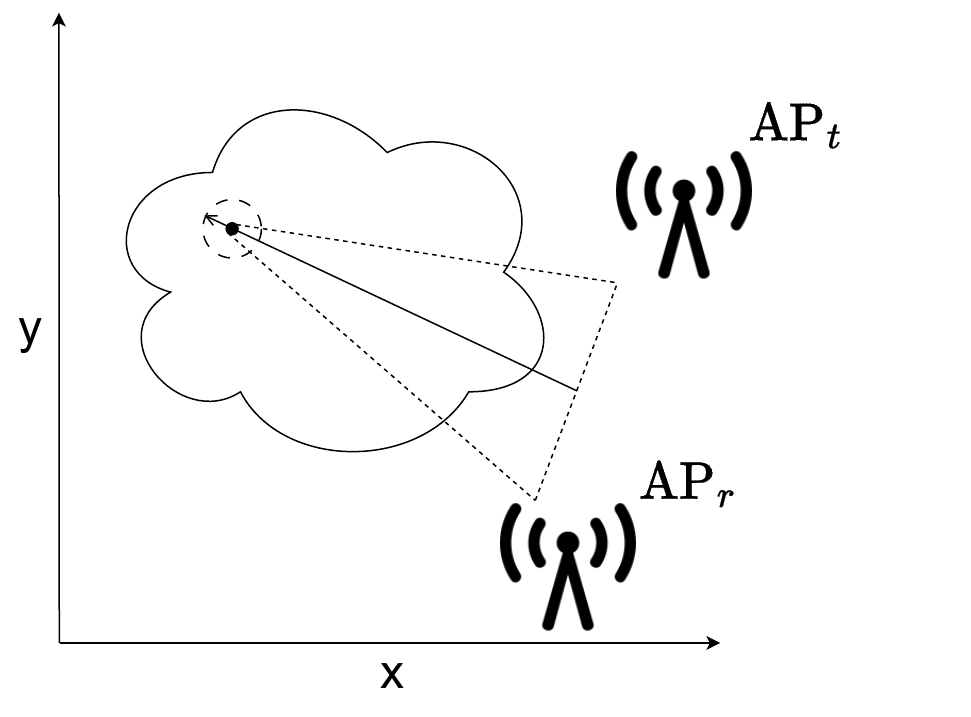}
	\vspace{-5pt}
	\caption{The minimum point is located on the boundary of the area.}
	\label{fig:edge}
	\vspace{-5pt}
\end{figure}

\vspace{-4pt}
For this problem, we apply grid search on the boundary points of the region $\mathcal{B}$. 
The minimum value of the objective function for any pair of APs occurs on the boundary points of the area, as depicted in Fig. \ref{fig:edge}. In this figure, by considering an interior point within the area, when this interior point is shifted perpendicularly to the line connecting AP $t$ and AP $r$, the distances from the point to both serving APs increase. Consequently, the point inevitably reaches a boundary point in this process.

Under these considerations, we propose the following algorithm to solve Problem $\mathcal{O} \mathcal{P}_{\text{CR}}$; the steps are given below:
\begin{enumerate}
	\item Select the AP that is closest to the centroid of the region $\mathcal{B}$, e.g., AP $t$.

	\item For the selected AP $t$, find $m_{r,t} = \underset{(x,y)}{\text{min}} \frac{1}{d_{r,(x,y)}^2 d_{t,(x,y)}^2}$ for each $r\ (r \neq t)$ by a grid search on the boundary points of the given area. Select the highest $\kappa$ values of $m_{r,t}$'s and save the indices $r$ of the selected highest $m_{r,t}$'s.
	
	\item Create a set of APs, i.e., set $\mathcal{S}$, with saved indices $r$ in the previous step and also add the selected CE in the first step to this set.
	For example, if we select AP 3 as a CE in Step 1 and find indices 12 and 18 for $\kappa=2$ in Step 2, then the set includes APs 3,12, and 18, i.e., $\mathcal{S}=\{3,12,18\}$.
	
	\item Calculate $m_{r,t} = \underset{(x,y)}{\text{min}} \frac{1}{d_{r,(x,y)}^2 d_{t,(x,y)}^2}$ by grid search on the boundary points for all remaining combinations of the elements of $\mathcal{S}$. For the given example, the possible combinations are $(r,t)=\{(12,3), (18,3), (12,18)\}$, and $m_{12,3}$ and $m_{18,3}$ are calculated in Step 2.
	
	\item Select the pair that gives the maximum $m_{r,t}$. The selected pair is our CE and reader.
\end{enumerate}
Due to the use of $\kappa$, we can determine the APs that are close to the region $\mathcal{B}$, and consequently decrease the computational effort. 
The proposed algorithm usually guarantees the optimal solution for a properly selected $\kappa$ value. The value of $\kappa$ could be chosen based on the number of APs, the size of the region $\mathcal{B}$, and the size of the coverage area where APs are located. For example, for a small value of $K$ and/or size of  the region $\mathcal{B}$, $\kappa$ could be chosen small.

\vspace{-3pt}
\section{Numerical Results}
\vspace{-5pt}
In this section, 
we evaluate the proposed algorithms for AP selection. To that purpose, we consider the following simulation parameters: $M=8, \gamma_1=1, \gamma_0=0,$ and $\kappa=2, 6$. The region $\mathcal{B}$ is selected as a square area.
Moreover, we select the learning rate in \gls{pgd} as $2000$, and the maximum number of iterations as $100$. The channels are modeled as free-space \gls{los}. A summary of the simulation parameters is provided in Table \ref{tab:simulationParameters}.

\begin{table}[tbp]
	\caption{Simulation Parameters}
	\vspace{-5pt}
	\centering
	\label{tab:simulationParameters}
	\resizebox{0.47\textwidth}{!}{\begin{tabular}{|l|r|}	
			\hline 
			\textbf{Parameter} &  \textbf{Value} \\ \hline\hline			
			Number of APs & $K=20, 30, 50$ \\ \hline
			Number of antennas per AP & $M=8$ \\ \hline
			Reflection coefficients of the BD & $\gamma_1=1, \gamma_0=0$ \\ \hline
			Size of the coverage area in meters & $30 \times 30, 40 \times 40$ \\ \hline
			Number of the highest  $m_{r,t}$ in Algorithm 2 & $\kappa=2, 6$ \\ \hline
			Size of the square region in meters & $5\times5, 10\times10$ \\ \hline			
			Learning rate in PGD & $2000$ \\ \hline
			Maximum number of iteration in PGD & $100$ \\
			\hline			
	\end{tabular}}
\vspace{-8pt}
\end{table}

\begin{figure}[tbp] 
	\centering
	\includegraphics[width =0.92\linewidth]{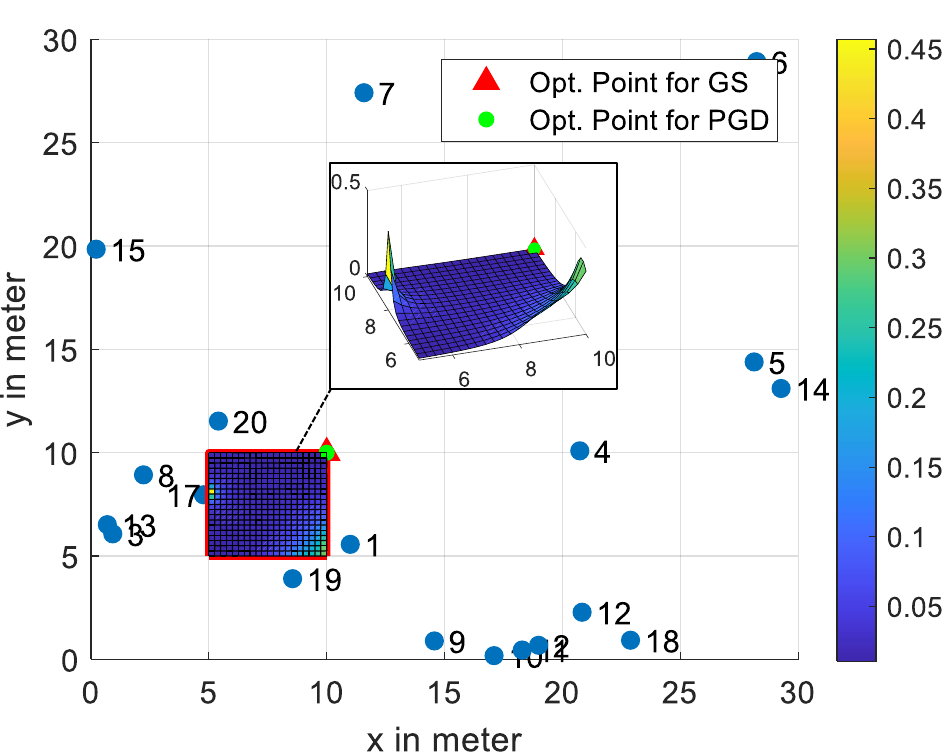}
\vspace{-5pt}
	\caption{Simulation result when AP 1 is CE and all the remaining APs are readers.}
\vspace{-8pt}
	\label{fig:APandBDarea}
\end{figure}

In Fig. \ref{fig:APandBDarea}, the AP locations and the square region where a BD is located are given. The number of APs is $K=20$, and the coverage area is $30 \times 30 \text{ m}^2$. The region $\mathcal{B}$ is a $5\times5 \text{ m}^2$ square area with the center at $(7.5,7.5)$ in meters. 
We solve the $\mathcal{O} \mathcal{P}_{\text{C}}$ to find the best CE, with all remaining APs are assigned as readers. 
Note that, although AP 17 is the closest AP to the centroid of the square area, it is not the optimal CE. We use \gls{gs} and \gls{pgd} to find the optimal CE, and both algorithms return AP 1 as the CE. It is worth mentioning that the computational complexity of \gls{pgd} is less than \gls{gs}.
In the region $\mathcal{B}$, the value of the objective function of $\mathcal{O} \mathcal{P}_{\text{C}}$ is shown when AP 1 is selected as the CE, and the optimal point shows the solution for $\mathcal{O} \mathcal{P}_{\text{C}1}$. As seen in Fig. \ref{fig:APandBDarea}, at $(x,y) = (10,10)$, we have the maximum $P_e$ when AP $1$ is CE.

\begin{figure}[tbp]
	\centering
	\includegraphics[width =0.84\linewidth]{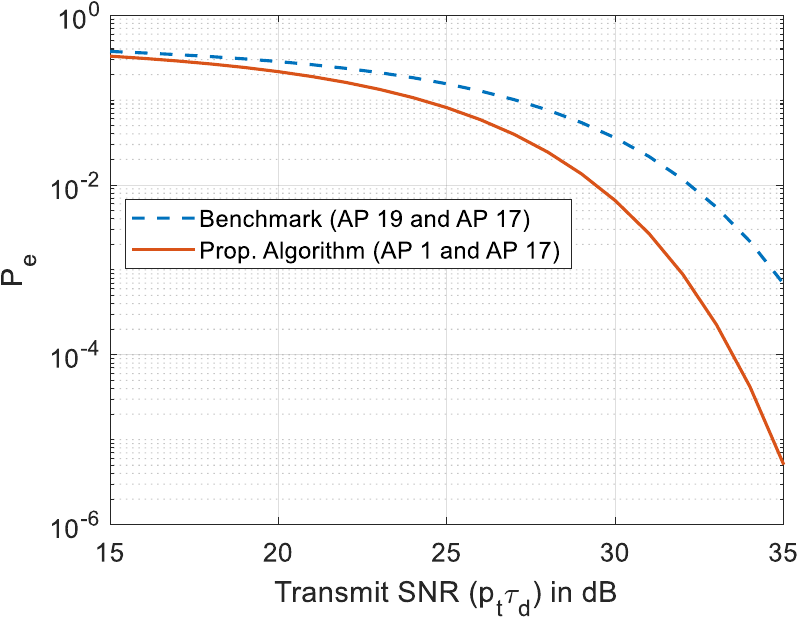}
	\vspace{-5pt}
	\caption{Comparison of the probability of error for the optimal CE-reader pair and the benchmark scenarios.}
	\label{fig:Pe}
	\vspace{-10pt}
\end{figure}

We also solve $\mathcal{O} \mathcal{P}_{\text{CR}}$ using the algorithm proposed in Section \ref{sec:bestCEreaderPairs} to find the best CE-reader pair for the network in Fig. \ref{fig:APandBDarea}. The algorithm gives AP 1 and AP 17 as the best CE and reader pair to serve the square region using the max-min fairness policy, and this result is the optimal solution. To establish a benchmark, we select the CE and reader pair that are the closest APs to the centroid of the square region, i.e., AP 17 and AP 19 in the deployment of Fig. \ref{fig:APandBDarea}. 
Fig. \ref{fig:Pe} shows the worst-case probability of error calculated as in Eq. \eqref{eq:P_e_Case1_and_2} versus the transmit SNR, for a point within region $\mathcal{B}$.
Note that for this setup, the minimum performance of the optimal solution outperforms the minimum performance of the benchmark scenario by $2.7$ dB at $P_e = 10^{-3}$.

\begin{figure}[tbp]
	\centering
	\includegraphics[width =0.89\linewidth]{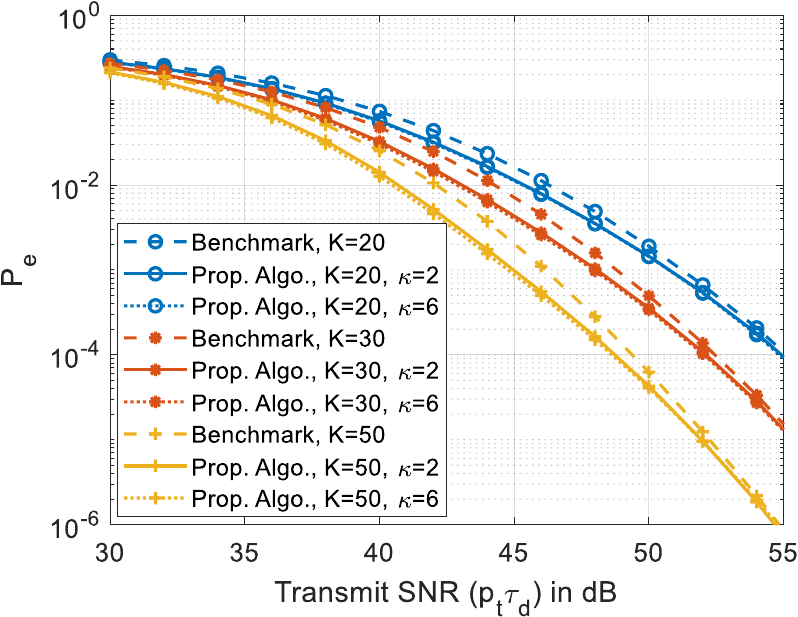}
	\vspace{-5pt}
	\caption{Comparison of the average probability of error for the optimal CE-reader pair and the benchmark scenarios.}
	\label{fig:PeMC}
	\vspace{-10pt}
\end{figure}

In Fig. \ref{fig:PeMC}, Monte-Carlo simulations are run to show the average performance improvement in $P_e$. For each iteration, both the locations of APs and the square area are deployed uniformly at random in the coverage area, and $P_e$ is calculated as in Eq. \eqref{eq:P_e_Case1_and_2}. The dimensions of the coverage area, both in width and length, are set as $40$ m, and the region $\mathcal{B}$ is defined as $10\times10 \text{ m}^2$. The number of APs is $K=20,30, \text{and }50$, and $\kappa=2$ and $6$. We use the algorithm proposed in Section \ref{sec:bestCEreaderPairs} to find the best CE-reader pair. As shown in Fig. \ref{fig:PeMC}, the error performance increases with the increasing number of APs, showing the advantage of cell-free \gls{mimo} network.
The proposed algorithm outperforms the benchmark by approximately $0.5$ dB, $0.9$ dB, and $1.4$ dB at $P_e = 10^{-3}$ for $K=20,30,$ and $50$, respectively. This improvement shows the effectiveness of our approach in minimizing $P_e$. In addition, the error performance of the proposed algorithm is almost the same for $\kappa=2$ and $\kappa=6$. Therefore, a selection of small $\kappa$ values is sufficient to achieve the optimal performance in this setup, and that decreases the computational complexity.

\vspace{-3pt}
\section{Conclusion}
\vspace{-3pt}
This work investigates \gls{ap} selection in \gls{bibc} operating in a cell-free \gls{mimo} network. We derive the \gls{map} detector to decode the \gls{bde} information bits in cell-free network, and additionally derive the closed-form expression of the probability of error for the detector.
We show that the performance of the detector remains unaffected regardless of the role assigned to each AP as a CE or reader in the case of two APs. However, when more than two APs are involved and these APs transmit orthogonal signals into the channel, we show that selecting a single AP as a CE and the remaining APs as readers is the optimal scenario to improve the \gls{bc} performance. 
We propose a solution to select the best possible CE to serve a region where \glspl{bde} can be located. We also propose a novel algorithm to select the best \gls{ce}-reader pair to serve the region. 
It is shown that the error performance of the proposed algorithm outperforms the error performance of the benchmark scenario. It is also shown that the cell-free \gls{mimo} network improves the performance of \gls{bc} by decreasing the round-trip path loss effect. 
As part of future work, we will investigate the AP selection algorithm with a beamformer and modified detector designs in the case of imperfect channel state information. The complexity analysis of PGD and GS will also be left for future work.

\appendices

\vspace{-5pt}
\section{The Probabilty of Error for the Case 3}
\vspace{-5pt}
\label{FirstAppendix}
The hypotheses are given as follows:
\vspace{-5pt}
\begin{equation} 
	\begin{split}
		\mathcal{H}_{0}&: \Y_{r}^l = \sum\limits_{t=t_1}^{t_T} \left( \G_{t, r} \matr{\Phi} c_t^l +\gamma_0 \g_{r} \g_{t}^\trp  \matr{\Phi} c_t^l \right) + \W_{r}^l,\\
		\mathcal{H}_{1}&: \Y_{r}^l =\sum\limits_{t=t_1}^{t_T} \left( \G_{t, r} \matr{\Phi} c_t^l +\gamma_1 \g_{r} \g_{t}^\trp  \matr{\Phi} c_t^l \right) + \W_{r}^l,
	\end{split}
	\vspace{-5pt}
\end{equation}
where $\Y_{r}^l$ and $\W_{r}^l$ denotes the received signal and noise, respectively, at the $r$$-$$\text{th}$ AP at time slot $l \in \{1,\ldots,T\}$, and the elements of $\W_{r}^l$ are i.i.d. $\mathcal{CN}(0, 1)$.
The value $c_t^l\ (|c_t^l|^2=\eta_t)$ stands for the $l$$-$th element of the orthogonal sequence used in AP $t$, and $\eta_t$ is the power control coefficient at the $t$$-$$\text{th}$ AP. 
When we follow similar calculations as in Section \ref{sec:info_det}, $P_e$ is
\vspace{-3pt}
\begin{equation} \label{eq:P_e_Case3}
	P_e = Q\left((\gamma_1-\gamma_0)\sqrt{\frac{T\eta_t}{2} \sum\limits_{t=t_1}^{t_T}\sum\limits_{\substack{r=r_1 \\ r\notin \{t_{1},\ldots, t_{T}\}}}^{r_R} ||\matr{A}_{t,r}||^2} \right),
\end{equation}
\vspace{-3pt}
\hspace{-8pt} where $\eta_t$ is selected as $\eta_t=1/T$ to satisfy the total radiated energy constraint.

\vspace{-3pt}
\section*{Acknowledgement}
\vspace{-3pt}
This work was funded by the REINDEER project of the European Union's Horizon 2020 research and innovation program under grant agreement No. 101013425, and in part by ELLIIT and the KAW foundation.
\vspace{-3pt}

\bibliographystyle{IEEEtran}
\bibliography{references}

\begin{thebibliography}{10}
\providecommand{\url}[1]{#1}
\csname url@samestyle\endcsname
\providecommand{\newblock}{\relax}
\providecommand{\bibinfo}[2]{#2}
\providecommand{\BIBentrySTDinterwordspacing}{\spaceskip=0pt\relax}
\providecommand{\BIBentryALTinterwordstretchfactor}{4}
\providecommand{\BIBentryALTinterwordspacing}{\spaceskip=\fontdimen2\font plus
\BIBentryALTinterwordstretchfactor\fontdimen3\font minus
  \fontdimen4\font\relax}
\providecommand{\BIBforeignlanguage}[2]{{%
\expandafter\ifx\csname l@#1\endcsname\relax
\typeout{** WARNING: IEEEtran.bst: No hyphenation pattern has been}%
\typeout{** loaded for the language `#1'. Using the pattern for}%
\typeout{** the default language instead.}%
\else
\language=\csname l@#1\endcsname
\fi
#2}}
\providecommand{\BIBdecl}{\relax}
\BIBdecl

\bibitem{3gpp.36.331}
\BIBentryALTinterwordspacing
``{3GPP TSG RAN Meeting -94e, Study proposal on passive IoT, 8A.1 (from
  RP-213368)},'' Tech. Rep., Dec. 2021. [Online]. Available:
  \url{https://www.3gpp.org/dynareport?code=TDocExMtg--RP-94-e--60214.htm}
\BIBentrySTDinterwordspacing

\bibitem{galappaththige2023cell}
D.~Galappaththige, F.~Rezaei, C.~Tellambura, and A.~Maaref, ``Cell-free
  bistatic backscatter communication: {C}hannel estimation, optimization, and
  performance analysis,'' \emph{arXiv preprint arXiv:2310.01264}, 2023.

\bibitem{kaplan2023direct}
A.~Kaplan, J.~Vieira, and E.~G. Larsson, ``Direct link interference suppression
  for bistatic backscatter communication in distributed {MIMO},'' \emph{IEEE
  Trans. Wireless Commun.}, early access, Jun. 2023.

\bibitem{mishra2019optimal}
D.~Mishra and E.~G. Larsson, ``{Optimal channel estimation for
  reciprocity-based backscattering with a full-duplex MIMO reader},''
  \emph{IEEE Trans. Signal Process.}, vol.~67, no.~6, pp. 1662--1677, Mar.
  2019.

\bibitem{kashyap2016feasibility}
S.~Kashyap, E.~Bj{\"o}rnson, and E.~G. Larsson, ``On the feasibility of
  wireless energy transfer using massive antenna arrays,'' \emph{IEEE Trans.
  Wireless Commun.}, vol.~15, no.~5, pp. 3466--3480, May 2016.

\bibitem{liu2014multi}
L.~Liu, R.~Zhang, and K.-C. Chua, ``Multi-antenna wireless powered
  communication with energy beamforming,'' \emph{IEEE Trans. Commun.}, vol.~62,
  no.~12, pp. 4349--4361, Dec. 2014.

\bibitem{kimionis2014increased}
J.~Kimionis, A.~Bletsas, and J.~N. Sahalos, ``Increased range bistatic scatter
  radio,'' \emph{IEEE Trans. Commun.}, vol.~62, no.~3, pp. 1091--1104, Mar.
  2014.

\bibitem{qu2022channel}
L.~Qu, D.~Mishra, and J.~Yuan, ``Channel estimation protocol for bistatic
  backscattering using multiantenna transceiver,'' in \emph{Proc. IEEE 33th
  Annu. Int. Symp. Pers., Indoor Mobile Radio Commun. (PIMRC)}, Dec. 2022, pp.
  439--444.

\bibitem{rezaei2023time}
F.~Rezaei, D.~Galappaththige, C.~Tellambura, and A.~Maaref, ``Time-spread
  pilot-based channel estimation for backscatter networks,'' \emph{arXiv
  preprint arXiv:2305.17248}, 2023.

\bibitem{van2018ambient}
N.~Van~Huynh, D.~T. Hoang, X.~Lu, D.~Niyato, P.~Wang, and D.~I. Kim, ``{Ambient
  backscatter communications: A contemporary survey},'' \emph{IEEE Commun.
  Surveys Tuts.}, vol.~20, no.~4, pp. 2889--2922, May. 2018.

\bibitem{liu2013ambient}
V.~Liu, A.~Parks, V.~Talla, S.~Gollakota, D.~Wetherall, and J.~R. Smith,
  ``{Ambient backscatter: Wireless communication out of thin air},'' in
  \emph{Proc. ACM SIGCOMM Conf.}, vol.~43, no.~4, Aug. 2013, pp. 39--50.

\bibitem{guo2018exploiting}
H.~Guo, Q.~Zhang, S.~Xiao, and Y.-C. Liang, ``Exploiting multiple antennas for
  cognitive ambient backscatter communication,'' \emph{IEEE Internet Things
  J.}, vol.~6, no.~1, pp. 765--775, Feb. 2019.

\bibitem{chen2022survey}
S.~Chen, J.~Zhang, J.~Zhang, E.~Bj{\"o}rnson, and B.~Ai, ``{A survey on
  user-centric cell-free massive MIMO systems},'' \emph{Digital Commun. Netw.},
  vol.~8, no.~5, pp. 695--719, Oct. 2022.

\bibitem{interdonato2019ubiquitous}
G.~Interdonato, E.~Bj{\"o}rnson, H.~Quoc~Ngo, P.~Frenger, and E.~G. Larsson,
  ``{Ubiquitous cell-free massive MIMO communications},'' \emph{EURASIP J.
  Wireless Commun. and Netw.}, vol. 2019, no.~1, Dec. 2019.

\bibitem{ammar2021user}
H.~A. Ammar, R.~Adve, S.~Shahbazpanahi, G.~Boudreau, and K.~V. Srinivas,
  ``{User-centric cell-free massive MIMO networks: A survey of opportunities,
  challenges and solutions},'' \emph{IEEE Commun. Surveys Tuts.}, vol.~24,
  no.~1, pp. 611--652, Dec. 2021.

\bibitem{demir2021foundations}
{\"O}.~T. Demir, E.~Bj{\"o}rnson, and L.~Sanguinetti, ``{Foundations of
  user-centric cell-free massive MIMO},'' \emph{FNT Signal Process.}, vol.~14,
  no. 3-4, pp. 162--472, Aug. 2021.

\bibitem{tse2005fundamentals}
D.~Tse and P.~Viswanath, \emph{{Fundamentals of Wireless Communication}}.\hskip
  1em plus 0.5em minus 0.4em\relax Cambridge, U.K.: Cambridge Univ. Press,
  2005.

\end{thebibliography}

\end{document}